\documentclass{emulateapj}

\newcommand{\msun}{\mbox{M$_{\odot}$}}
\newcommand{\kms}{\mbox{km s$^{-1}$}}

\newcommand{\gps}{\ensuremath{g_{\rm P1}}}
\newcommand{\rps}{\ensuremath{r_{\rm P1}}}
\newcommand{\ips}{\ensuremath{i_{\rm P1}}}
\newcommand{\zps}{\ensuremath{z_{\rm P1}}}
\newcommand{\yps}{\ensuremath{y_{\rm P1}}}

\slugcomment{Submitted to Astrophysical Journal Letters}

\shorttitle{SN 2011ht}
\shortauthors{Fraser et al.}

\begin{document}

\title{Detection of an outburst one year prior to the explosion of SN 2011ht}

\author{M. Fraser\altaffilmark{1},
	M. Magee\altaffilmark{1},
	R. Kotak\altaffilmark{1},
	S.J. Smartt\altaffilmark{1},
	K.W. Smith\altaffilmark{1},
	J. Polshaw\altaffilmark{1},
	A.J. Drake\altaffilmark{2},
	T. Boles\altaffilmark{3},
	C.-H. Lee\altaffilmark{4,5},
	W.S. Burgett\altaffilmark{6},  
	K.C. Chambers\altaffilmark{6},   
	P.W. Draper\altaffilmark{7},  
	H. Flewelling\altaffilmark{6}, 
	K.W. Hodapp\altaffilmark{6},   
	N. Kaiser\altaffilmark{6},    
	R.-P. Kudritzki\altaffilmark{6}, \\   
	E.A. Magnier\altaffilmark{6},  
	P.A. Price\altaffilmark{8}, 
	J.L. Tonry\altaffilmark{6},    
	R.J. Wainscoat\altaffilmark{6},  
	C. Waters\altaffilmark{6}.
	}

\altaffiltext{1}{School of Mathematics and Physics, Queens University Belfast, Belfast BT7 1NN; m.fraser@qub.ac.uk}
\altaffiltext{2}{Department of Astronomy and the Center for Advanced Computing Research, California Institute of Technology, Pasadena, CA  91225, USA}
\altaffiltext{3}{Coddenham Astronomical Observatory, Suffolk, UK}
\altaffiltext{4}{University Observatory Munich, Scheinerstrasse 1, 81679 Munich, Germany}
\altaffiltext{5}{Max Planck Institute for Extraterrestrial Physics, Giessenbachstrasse, 85748 Garching, Germany}
\altaffiltext{6}{Institute for Astronomy, University of Hawaii at Manoa, Honolulu, HI 96822, USA}
\altaffiltext{7}{Department of Physics, Durham University, South Road, Durham DH1 3LE, UK}
\altaffiltext{8}{Department of Astrophysical Sciences, Princeton University, Princeton, NJ 08544, USA}

\begin{abstract}
Using imaging from the Pan-STARRS1 survey, we identify a precursor outburst at 287 and 170 days prior to the 
reported explosion of the purported Type IIn supernova (SN) 2011ht. In the Pan-STARRS data, a source coincident with SN 2011ht is detected exclusively in the \zps\ and \yps-bands. An absolute magnitude of  M$_z\simeq$-11.8 suggests that this was an outburst of the progenitor star. 
Unfiltered, archival Catalina Real Time Transient survey images also reveal a coincident source from at least 258 to 138 days before the main event. We suggest that the outburst is likely to be  an intrinsically red eruption, although we cannot conclusively exclude a series of erratic outbursts which were observed only in the redder bands by chance. This is only the fourth detection of an outburst prior to a claimed SN, and lends credence to the possibility that many more interacting transients have pre-explosion outbursts, which have been missed by current surveys.
\end{abstract}


\section{Introduction}

Core-collapse supernovae (CCSNe) result from the death of a massive star that has exhausted its nuclear fuel, and can no longer resist gravitational collapse. Despite this common explanation, CCSNe show considerable observational diversity, resulting from the structure and composition of the progenitor star immediately prior to collapse \citep[][and references therein]{Sma09}. The observed properties of CCSNe are also influenced by their environments; the specific subclass of events, dubbed Type IIn SNe, are defined by the interaction between the SN ejecta and a surrounding region of dense circumstellar material (CSM) which typically gives rise to narrow (few $\sim$100s \kms) Balmer emission lines \citep{Sch90}. In many cases, the luminosity of Type IIn SNe can be enhanced many times over that of a normal core-collapse event due to the transfer of some fraction of the kinetic energy from the SN ejecta as it collides with the CSM. 

Given that mass loss is a key driver of massive star evolution, the most likely origin of the dense circumstellar material close to the SN would be from the progenitor star. This temporal coincidence suggests that the mass loss may be linked to the final evolutionary stages of the star, although the detailed physics of this process remains elusive. Some Type IIn SNe have been linked to very massive Luminous Blue Variable (LBV) stars \citep[e.g][]{Tru08,Gal09}, which are known to experience very high mass loss rates \citep{Hum94}. This poses a challenge for the current paradigm of stellar evolution theory, which would not lead us to expect LBVs to be the direct antecedent of CCSNe \citep{kv06}. Another intriguing connection is with the so-called SN impostors \citep{Van00}. SN impostors are observationally similar to Type IIn SNe, but the substantial difference in peak brightness compared to a {\it bona fide} CCSN, means that they are commonly thought to be (giant) eruptions of a massive star, where up to a few \msun\ of material can be ejected by the star, but {\it without undergoing core-collapse} \citep[e.g][]{Pas10}. The recent SN~2009ip \citep{Smi10,Fol11} has further complicated this picture, with a three year long series of outbursts, that culminated in an event which some authors have claimed is the star exploding as a CCSN, and others have suggested may be an extreme non-terminal eruption \citep{Pas13,Mau13a,Fra13,Mar13,Smi13a, Pri13}. Similar controversy also surrounds the fate of SN 1961V \citep{Zwi64,Koc11,Smi11a,Van12}. Aside from these debated transients, only two other SNe have purported eruptions prior to their final core-collapse; the Type Ibn SN 2006jc \citep{Pas07,Fol07} and the Type IIn SN 2010mc \citep{Ofe13}.

It is clear that a critical aspect of understanding Type IIn SNe is to identify and characterize the eruptions and outbursts which eject material into the CSM prior to the final core-collapse. To this end, we present in this Letter an analysis of the eruptive history of the Type IIn SN 2011ht, using serendipitous archival imaging taken prior to its discovery on 29 September 2011 (MJD 55833.68). When SN 2011ht was first discovered \citep{Bol11}, it was initially classified as a SN impostor \citep{Pas11}. However, the transient reached a peak absolute magnitude of $M_V\sim$-17, and \cite{Pri11} later claimed that it was in fact a Type IIn SN. The ambiguity as to the nature of SN 2011ht has continued, with \cite{Rom12} concluding that SN 2011ht had features reminiscent of both Type IIn SNe and SN impostors, although the observed UV emission had not been seen in a SN impostor before. Subsequently, \cite{Hum12} argued that SN 2011ht was {\it not} a CCSN, based on velocities, and estimates of energy and ejected mass. Most recently, \cite{Mau13b} presented late time (+147 d) spectra of SN 2011ht with weak nebular [O~{\sc ii}] emission; this indicator of nucleosynthetic processing favours a core-collapse interpretation.

In what follows, we adopt a kinematic distance of 19.2 Mpc towards UGC 5460, and a foreground extinction of $A_r$=0.025 \citep{Rom12}. The explosion epoch of SN 2011ht is uncertain, but is probably close to the discovery epoch based on the rising lightcurve and UV magnitudes. We therefore adopt the discovery epoch, 2011 Sept. 29; MJD 55833.68, as t$_0$. 

\section{Observations and data analysis}

The Panoramic Survey Telescope and Rapid Response System 1 (Pan-STARRS1; henceforth abbreviated to PS1) is a wide-field, panchromatic survey telescope located on Haleakala in Hawaii \citep{Kai10}. As part of the ``3$\pi$ Survey'', PS1 observes the entire visible night sky on a rolling basis in \gps\rps\ips\zps\yps\ filters \citep{Stu10}. These images typically reach a limiting magnitude of $\sim$20-21 mag, and hence are sufficiently deep to detect relatively faint precursor outbursts of nearby SNe. In the 3$\pi$ survey, two exposures separated by $\sim$30 min (termed a ``Transient Time Interval (TTI) pair'') are taken with the same filter and at an identical pointing for each epoch/visit; this permits the removal of moving objects and cosmic ray hits.  The images are reduced by the Image Processing Pipeline \citep[IPP;][]{Mag06,Mag07}, which detrends the data, performs astrometric and photometric calibration, stacks and template-subtracts images, and finally, performs source detection and photometry of all objects in the field.

We examined all available PS1 images of the site of SN 2011ht, from the earliest image taken in Feb 2010 until the most recent image from April 2013. A source was visible at the SN position in several frames; and for each of these we list the magnitude determined using point spread function (PSF) fitting photometry as implemented in IPP, in Table \ref{table_phot}. In all cases, the source was visible in both frames taken as part of the TTI pair. These magnitudes are in the PS1 photometric system as described in \cite{Ton12}, which is comparable to the SDSS photometric system for {\it z}-band ($|${\it z}$_\mathrm{SDSS}$-{\it z$_\mathrm{P1}$}$|\lesssim$0.05 mag). We additionally checked the IPP photometry for the {\it z}-band detection using independent PSF-fitting routines within the 
{\sc iraf}\footnote{{\sc iraf} is distributed by the National Optical Astronomy Observatory, which is operated by the Association of Universities for Research in Astronomy (AURA) under cooperative agreement with the National Science Foundation.} environment; a photometric zeropoint for each image was determined from aperture photometry of nearby sources with catalogued SDSS DR6 \citep{Ade08} magnitudes. Uncertainties were estimated using artificial star tests near the SN location, and added in quadrature with the uncertainty in the zeropoint. We found that both methods yielded magnitudes consistent to within 0.06 mag. For frames where a source was not visible at the position of SN 2011ht, we estimated a limiting magnitude (typically $\gtrsim$20--21) by adding progressively fainter artificial sources to the image, using a PSF built from nearby point sources. The limiting magnitude was taken to be that of the faintest artificial source which could be clearly identified by visual inspection at the position of SN 2011ht. We note that this is a conservative approach; photometry on the faintest recovered artificial sources has a typical error of $\sim$0.1 mag, corresponding to a $\sim10 \sigma$ detection.

\begin{deluxetable*}{lllcccccr} 
\tablecolumns{9} 
\tablewidth{0pc} 
\tablecaption{PS1 magnitudes for SN 2011ht, with respect to t$_0$ = 55833.68. As we have used CSS primarily for pre-September 2011 data, we do not list the CSS photometry of SN 2011ht here. Pre- and post- SN 2011ht photometry are separated with a line. We note that the unfiltered CRTS images (marked with an asterisk) are calibrated to SDSS {\it r}.\label{table_phot}} 
\tablehead{ 
\colhead{Date (UT)} & \colhead{MJD}   & \colhead{Phase}    & \colhead{g (err)} & \colhead{r (err)}    & \colhead{i (err)}   & \colhead{z (err)}    & \colhead{y (err)}	& \colhead{Instrument} }
\startdata 
2009 Nov 06 	& 55141.470	& -692.21	& - 				& $>$20.3	*		& -				& -				& -				& CSS		\\
2009 Nov 20 	& 55155.460	& -678.22	& - 				& $>$20.2	*		& -				& -				& -				& CSS		\\
2009 Dec 18 	& 55183.381	& -650.30	& - 				& $>$20.1	*		& -				& -				& -				& CSS		\\
2010 Feb 03	& 55230.538	& -603.14	& -				& -				& $>$21.1			& -				& -				& PS1		\\
2010 Feb 03	& 55230.568	& -603.11	& -				& -				& $>$20.5			& -				& -				& PS1		\\
2010 Feb 20	& 55247.179   	& -586.50	& - 				& $>$20.2	*		& -				& -				& -				& CSS		\\
2010 Feb 22	& 55249.366	& -584.31	& -				& $>$21.1			& -				& -				& -				& PS1		\\
2010 Feb 22	& 55249.376	& -584.30	& -				& $>$21.2			& -				& -				& -				& PS1		\\
2010 Feb 25	& 55252.355	& -581.33	& -				& -				& $>$20.6			& -				& -				& PS1		\\
2010 Feb 25	& 55252.365	& -581.32	& -				& -				& $>$20.1			& -				& -				& PS1		\\
2010 Mar 18	& 55273.280   	& -560.40	& - 				& $>$20.1	*		& -				& -				& -				& CSS		\\
2010 Apr 12	& 55298.249   	& -535.43	& - 				& $>$20.3	*		& -				& -				& -				& CSS		\\
2010 May 11	& 55327.237   	& -506.44	& - 				& $>$20.1	*		& -				& -				& -				& CSS		\\
2010 Dec 16	& 55546.593	& -287.09	& -				& -				& -			 	& 19.597 (0.049) 	& -				& PS1		\\
2010 Dec 16	& 55546.605	& -287.09 & -				& -				& - 				& 19.702 (0.052)	& -				& PS1		\\
2011 Jan 14	& 55575.268	& -258.41	& -				& 20.32 (0.18)*		& - 				& -				& -				& CSS		\\ 
2011 Feb 10	& 55602.298	& -231.38	& -		        		& 20.37 (0.16)*		& - 				& -				& -				& CSS		\\ 
2011 Feb 23	& 55615.325	& -218.36	& $>$21.1			& -				& -				& -				& -				& PS1		\\
2011 Feb 23	& 55615.337	& -218.34	& $>$21.7			& -				& -				& -				& -				& PS1		\\
2011 Mar 12	& 55632.309 	& -207.37 & -				& 20.26 (0.17)*		& - 				& -				& -				& CSS		\\
2011 Apr 12	& 55663.271  	& -170.41	& -				& -				& -			 	& -				& 19.209 (0.145)	& PS1		\\
2011 May 13	& 55694.235 	& -139.45	& - 				& 20.59 (0.48)*		& - 				& -				& -				& CSS		\\
\hline
2012 Feb 13	& 55970.543	& 136.86	& -				& -				& 18.839 (0.019) 	& -				& -				& PS1		\\
2012 Feb 13	& 55970.555	& 136.88	& -				& -				& 18.852 (0.020) 	& -				& -				& PS1		\\
2012 Feb 26	& 55983.462	& 149.78	& 19.886 (0.030) 	& -				& -    				& -				& -				& PS1		\\
2012 Feb 26	& 55983.474	& 149.79	& 19.956 (0.032)	& -				& -				& -				& -				& PS1		\\
2012 Apr 01	& 56018.298	& 184.62	& -				& -				& -				& -				& 19.435 (0.147)	& PS1		\\
2012 Apr 01	& 56018.305	& 184.63	& -				& -				& -			 	& -				& 19.612 (0.152)	& PS1		\\
2012 Apr 02	& 56019.286	& 185.61	& -				& -				& -			 	& 19.573 (0.049) 	& -				& PS1		\\
2012 Apr 02	& 56019.293	& 185.61	& -				& -				& -			 	& 19.531 (0.052)	& -				& PS1		\\
2012 Dec 27	& 56288.590	& 454.90	& -				& $>$21.0			& -				& -				& -				& PS1		\\
2012 Dec 27	& 56288.601	& 454.92	& -				& $>$20.6			& -				& -				& -				& PS1		\\
2013 Jan 11	& 56303.561	& 469.88	& $>$21.4			& -				& -				& -				& -				& PS1		\\
2013 Jan 11	& 56303.573	& 469.89	& $>$21.4			& -				& -				& -				& -				& PS1		\\
2013 Jan 25	& 56317.529	& 483.85	& -				& -				& -				& -				& $>$19.6			& PS1		\\
2013 Jan 25	& 56317.536	& 483.86	& -				& -				& -				& -				& $>$20.2			& PS1		\\
2013 Jan 25	& 56317.537	& 483.86	& -				& -				& -				& $>$20.1			& -				& PS1		\\
2013 Jan 25	& 56317.539	& 483.86	& -				& -				& -				& $>$20.3			& -				& PS1		\\
2013 Feb 02	& 56325.448	& 491.77	& -				& $>$19.9			& -				& -				& -				& PS1		\\
2013 Feb 02	& 56325.460	& 491.78	& -				& $>$20.6			& -				& -				& -				& PS1		\\
2013 Feb 02	& 56325.472	& 491.79	& -				& -				& $>$20.4			& -				& -				& PS1		\\
2013 Feb 02	& 56325.485	& 491.81	& -				& -				& $>$21.1			& -				& -				& PS1		\\
2013 Feb 08	& 56331.440	& 497.76	& $>$21.0			& -				& -				& -				& -				& PS1		\\
2013 Feb 08	& 56331.453	& 497.77	& $>$21.1			& -				& -				& -				& -				& PS1		\\
2013 Apr 21	& 56403.273	& 569.59	& -				& -				& -				& -				& $>$19.8			& PS1		\\
2013 Apr 21	& 56403.284	& 569.60	& -				& -				& -				& -				& $>$19.7			& PS1		\\
2013 Apr 22	& 56404.245	& 570.57	& -				& -				& -				& $>$20.1			& -				& PS1		\\
2013 Apr 22	& 56404.252	& 570.57	& -				& -				& -				& $>$19.9			& -				& PS1		\\
\enddata
\end{deluxetable*}

\begin{figure*}
\centering
	\includegraphics[width=0.245\linewidth]{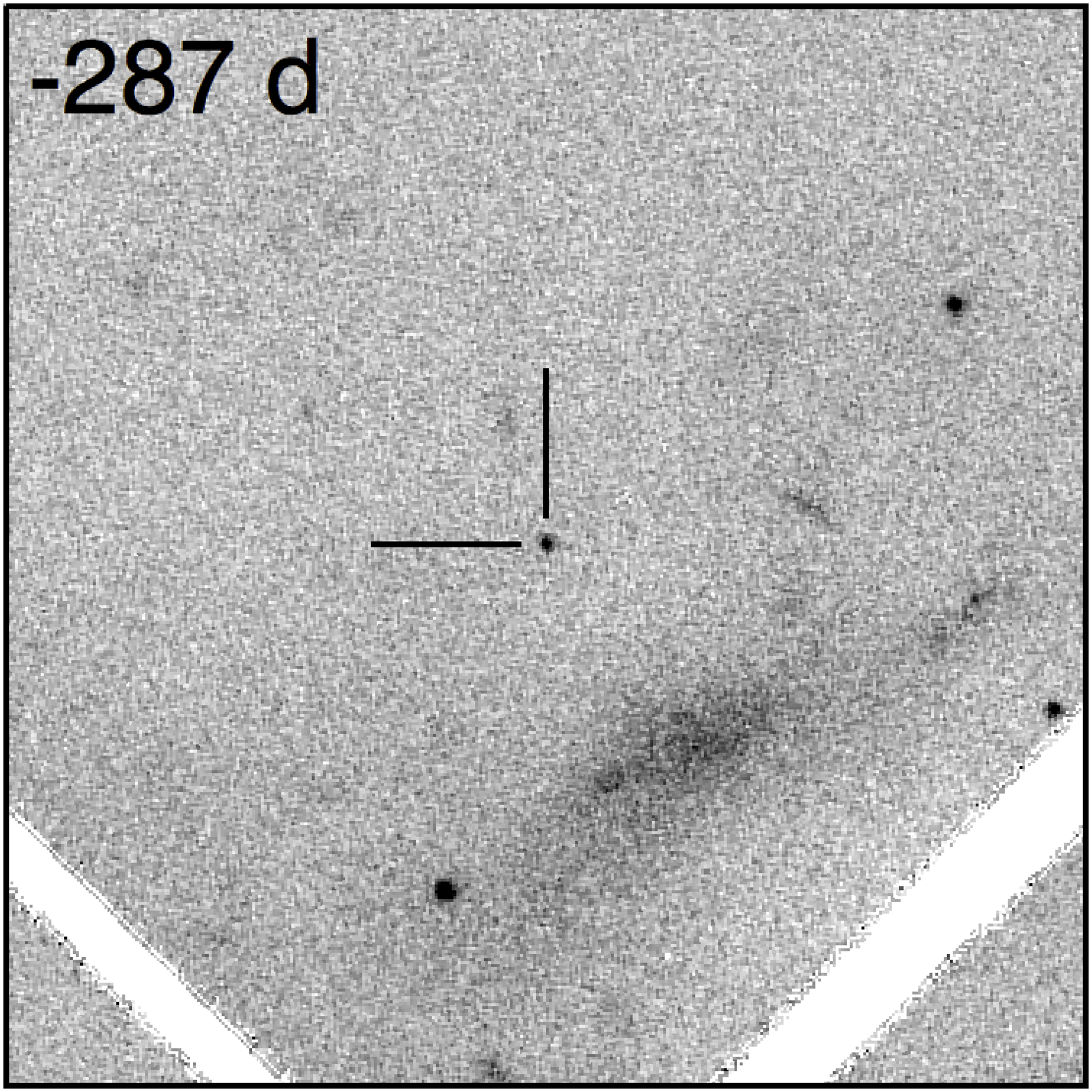}	
	\includegraphics[width=0.245\linewidth]{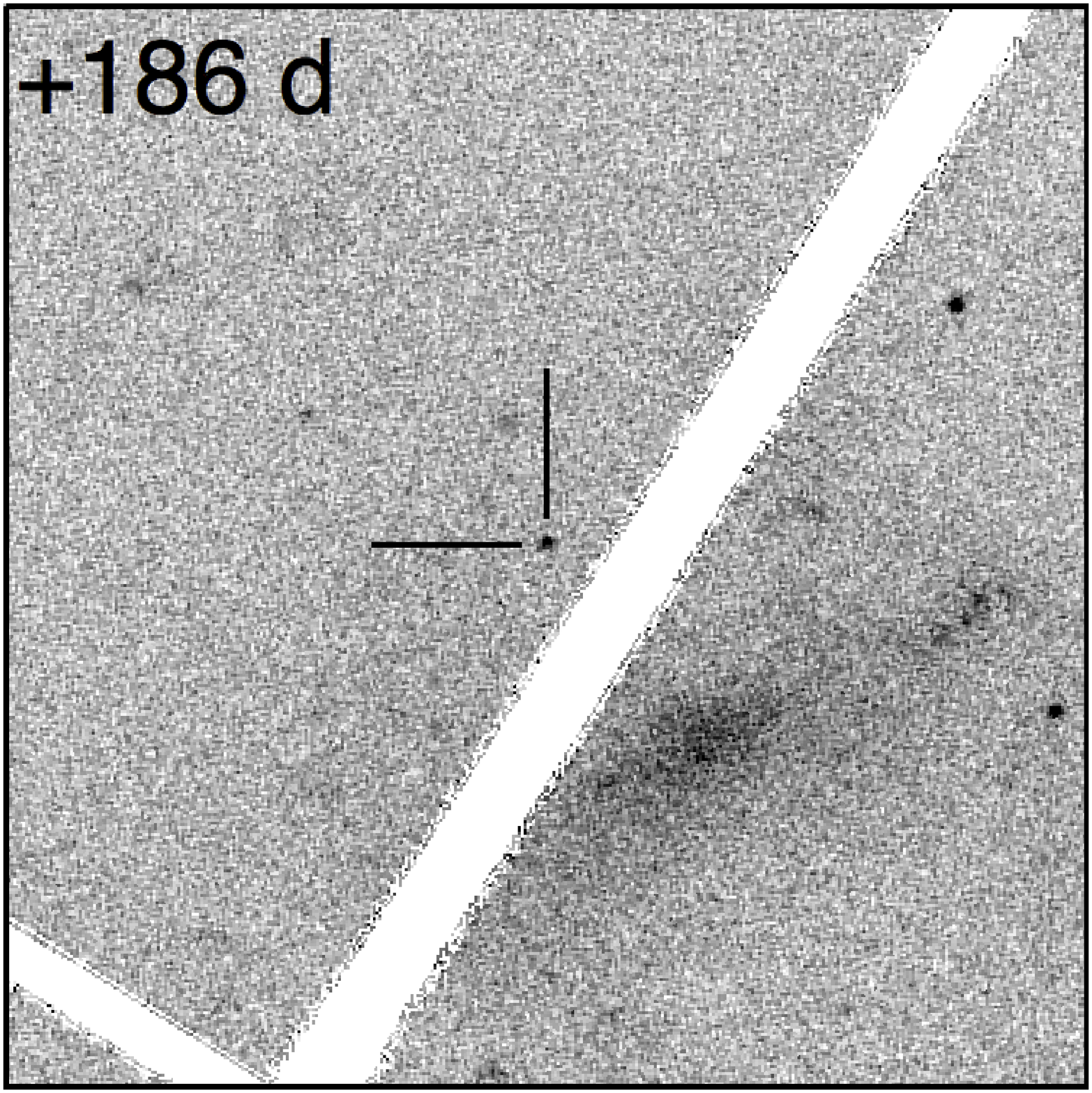}
	\includegraphics[width=0.245\linewidth]{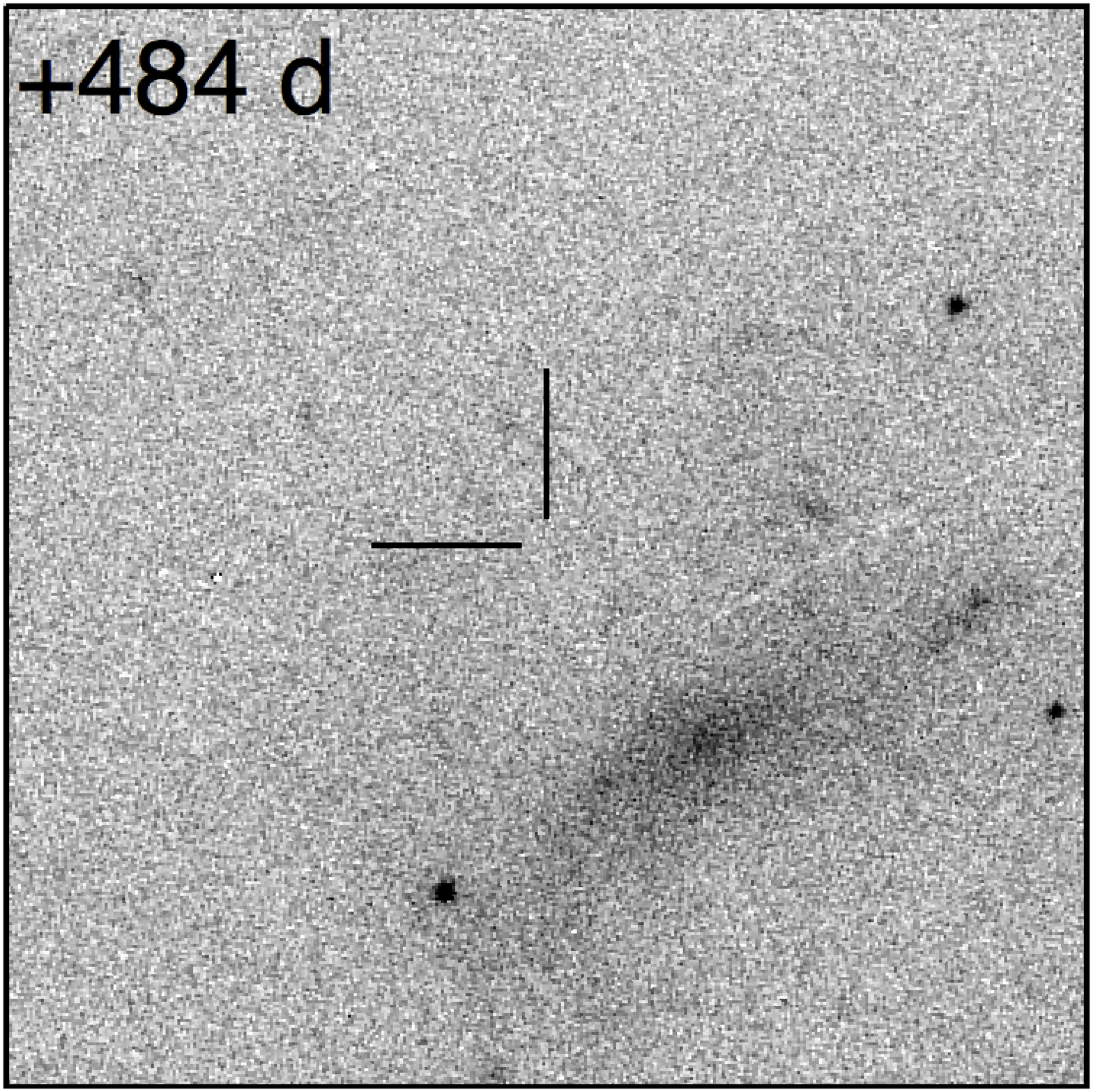}
	\includegraphics[width=0.245\linewidth]{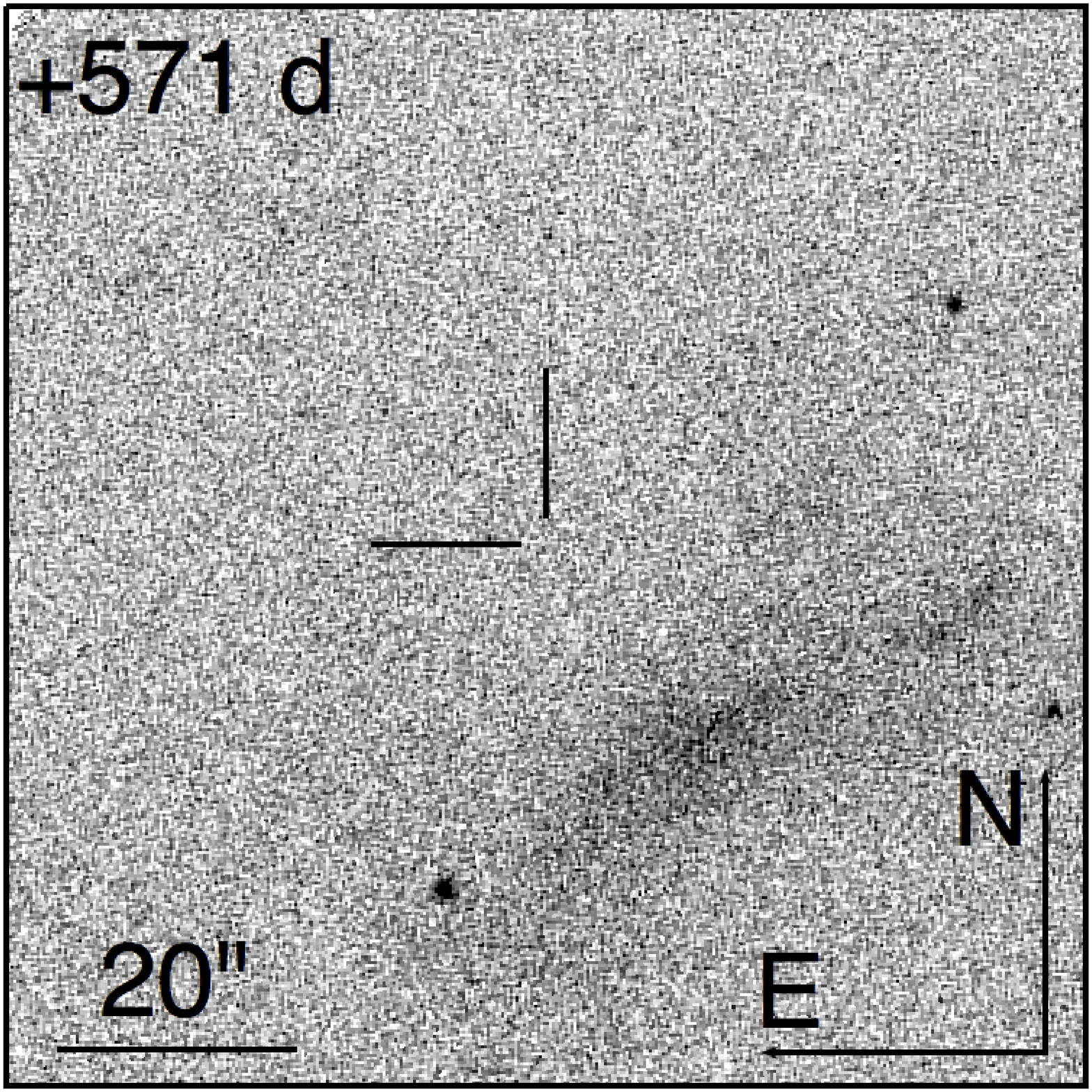}
\caption[]{PS1-{\it z}-band imaging of the site of SN 2011ht. The epochs indicated are with respect to the discovery epoch. The SN position is marked at the centre of the field. The white diagonal strips are gaps between the PS1 CCDs.}
\label{fig:z}
\end{figure*}

We clearly detect a point source coincident with SN 2011ht in {\it z}- and {\it y}-bands, as shown in Fig.~\ref{fig:z}, 287 and 170 d prior to the discovery of SN 2011ht in September 2011 respectively. We designate the source PSO J152.0441+51.8492. The PS1 images used are astrometrically calibrated by IPP against the 2MASS catalog \citep{Sku06}, and so we measured the pixel coordinates of PSO J152.0441+51.8492 in the {\it z}-band image from 2010 December 16, and compared to the pixel coordinates of SN 2011ht as measured in the two {\it z} images from 2012 April 2. We find the position of the source in each frame is coincident to 0.45 pixels (0.11\arcsec). To check the alignment of the images, we measured the positions of 11 other sources in the field of view which had comparable magnitudes. We found a mean random offset of 0.52 pixels for these sources, consistent with that measured for SN 2011ht / PSO J152.0441+51.8492. We hence conclude that the two sources are coincident. The measured magnitudes for PSO J152.0441+51.8492 are $\sim$1-2 mag brighter than the limiting magnitudes of the archival SDSS images in which \cite{Rom12} did not detect a progenitor for SN 2011ht, indicating that this is an outburst rather than a detection of the quiescent progenitor.

Unfortunately, there are no pre-explosion images of the site of SN 2011ht from the Spitzer Space Telescope. We examined the All-Sky Source Catalog produced by the Wide-field Infrared Survey Explorer \cite[WISE;][]{Wri10} for any source coincident with SN 2011ht at any epoch. No source was listed in the catalog for the epoch of the WISE observations (2010 April 29). WISE subsequently re-observed the site of SN 2011ht on 2010 November 6 during the WISE ``Post-Cyro'' phase. As the telescope had depleted its reserve of coolant, the post-cryo images are less sensitive than those taken in 2010. A low S/N ($<3 \sigma$) detection was reported within 2\arcsec\ of SN 2011ht in the post-cryo catalog, and so we downloaded the individual ``L1b'' images in the W1 and W2 filters from the Infrared Processing and Analysis Center archive. After aligning and stacking all post-cryo frames taken in each filter, we subtracted these from templates created from the 2010 images. No source could be seen in the difference image, leading us to conclude the detection in the post-cryo catalog was spurious. The WISE data does not yield particularly useful constraints on SN 20011ht, as the All-Sky Source Catalog is only $>$95 \% complete to W1=16.6 and W2=16.0. The low spatial resolution of WISE (6\arcsec) is less relevant, as SN 2011ht is not in a crowded region.

\begin{figure*}
\centering
	\includegraphics[width=0.245\linewidth]{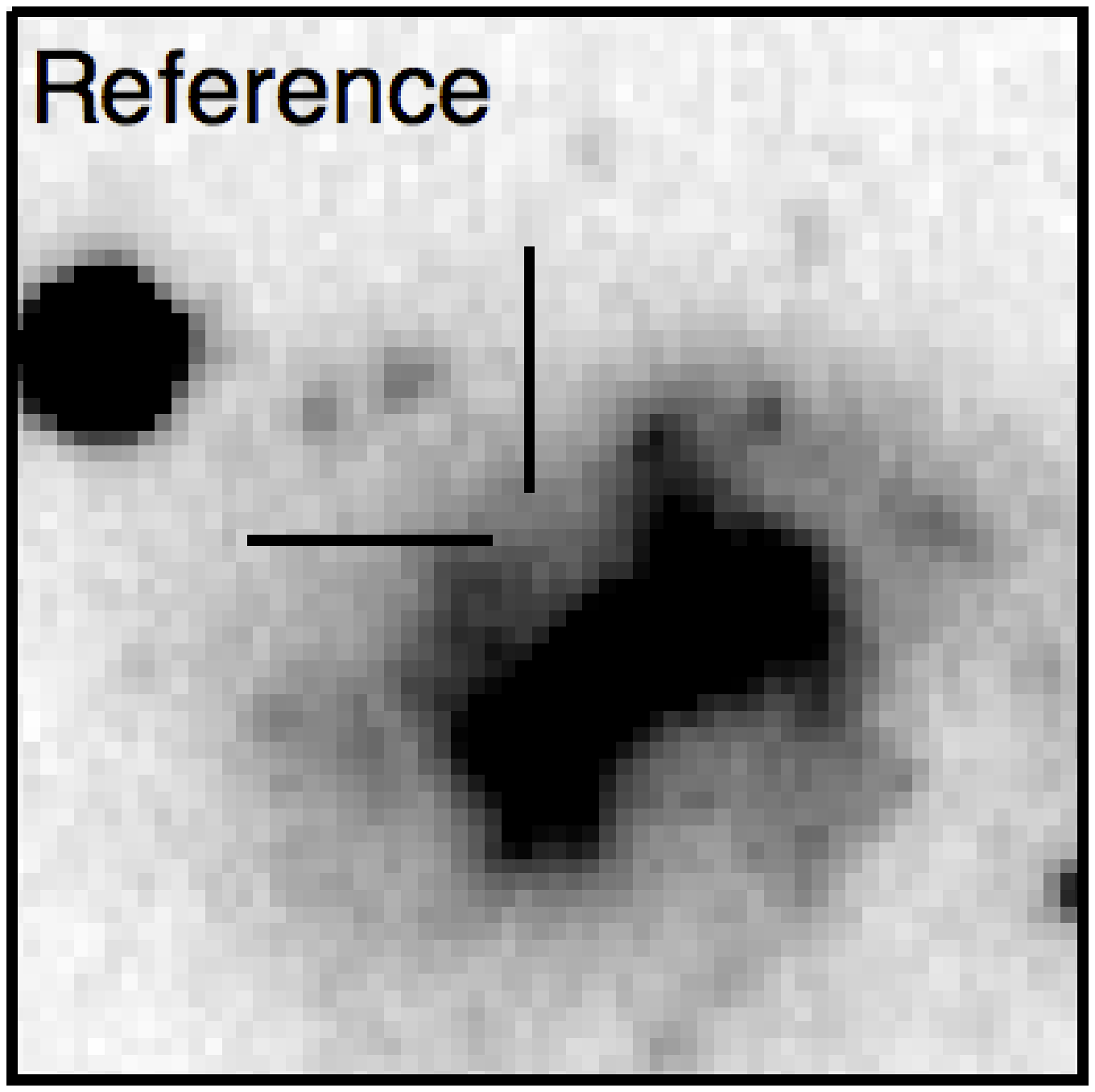}	
	\includegraphics[width=0.245\linewidth]{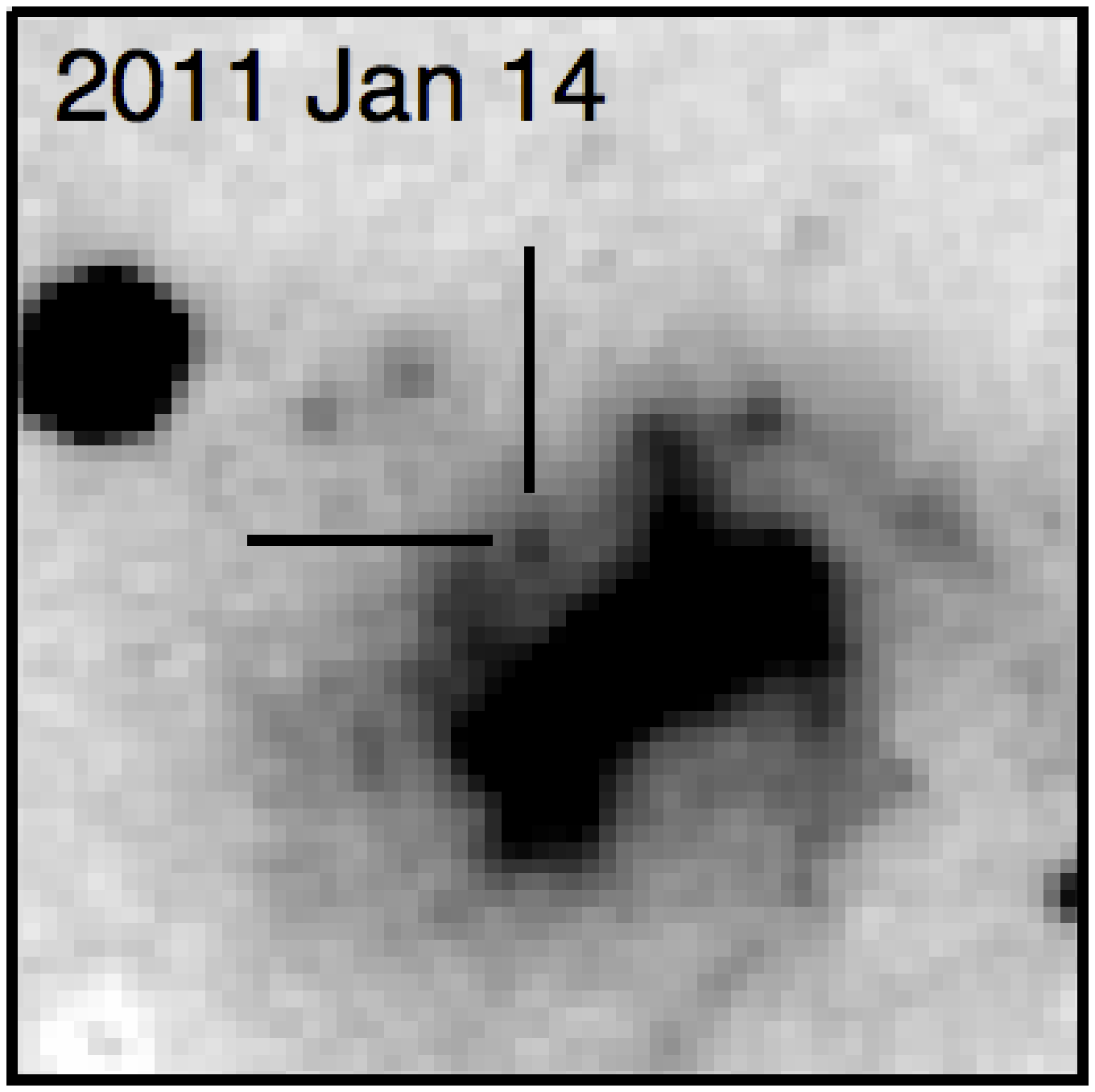}
	\includegraphics[width=0.245\linewidth]{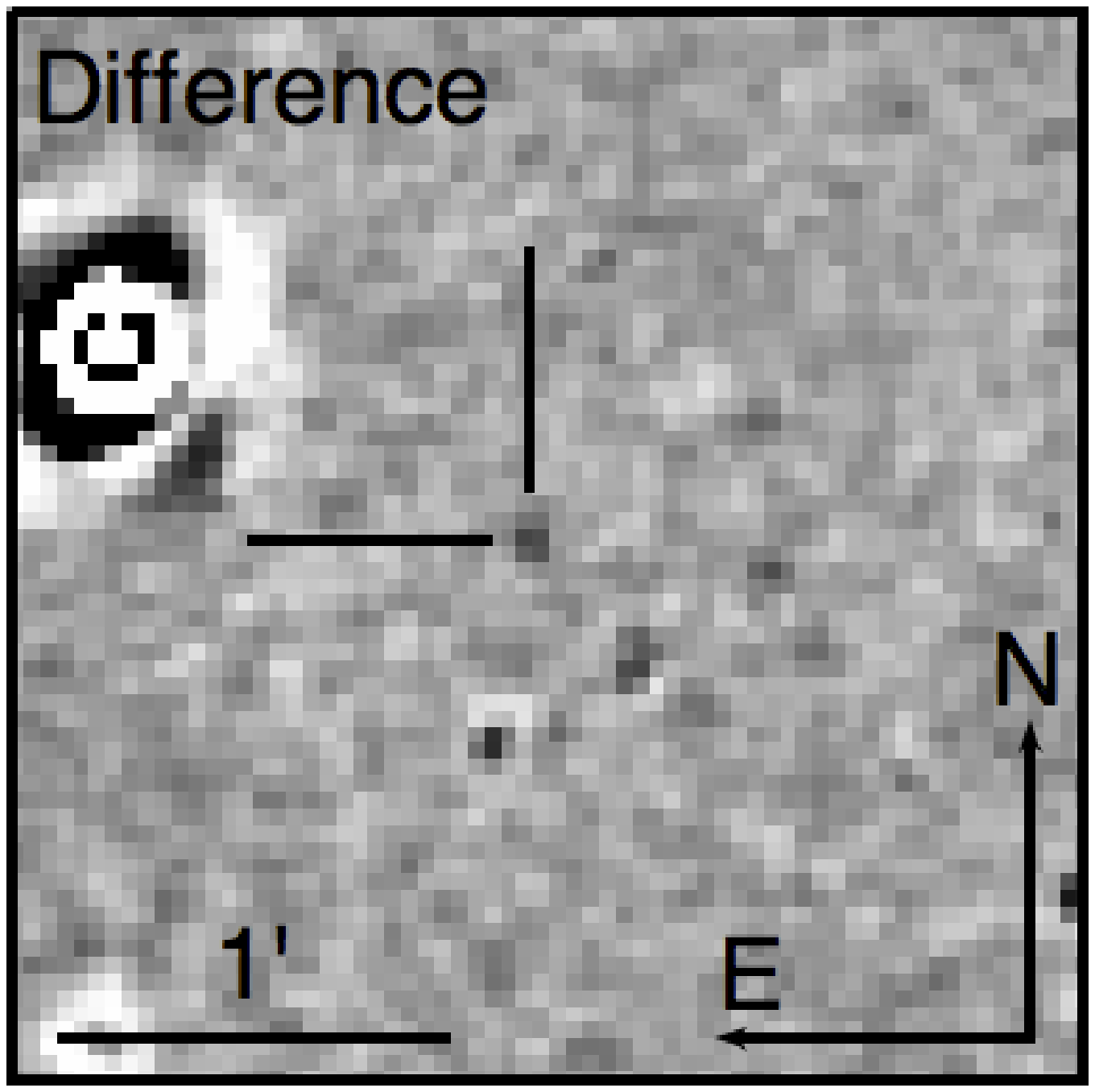}
\caption[]{Example of the subtractions obtained between CRTS images. The position of PSO J152.0441+51.8492 is indicated with tick marks in all frames. Some apparent faint sources besides PSO J152.0441+51.8492 can be seen in the difference image, these are artefacts due to an imperfect subtraction in the galaxy core. The bright star to the NW was saturated, leaving a residual.}
\label{fig:crts}
\end{figure*}

\begin{figure*}
\epsscale{1.10}
\plotone{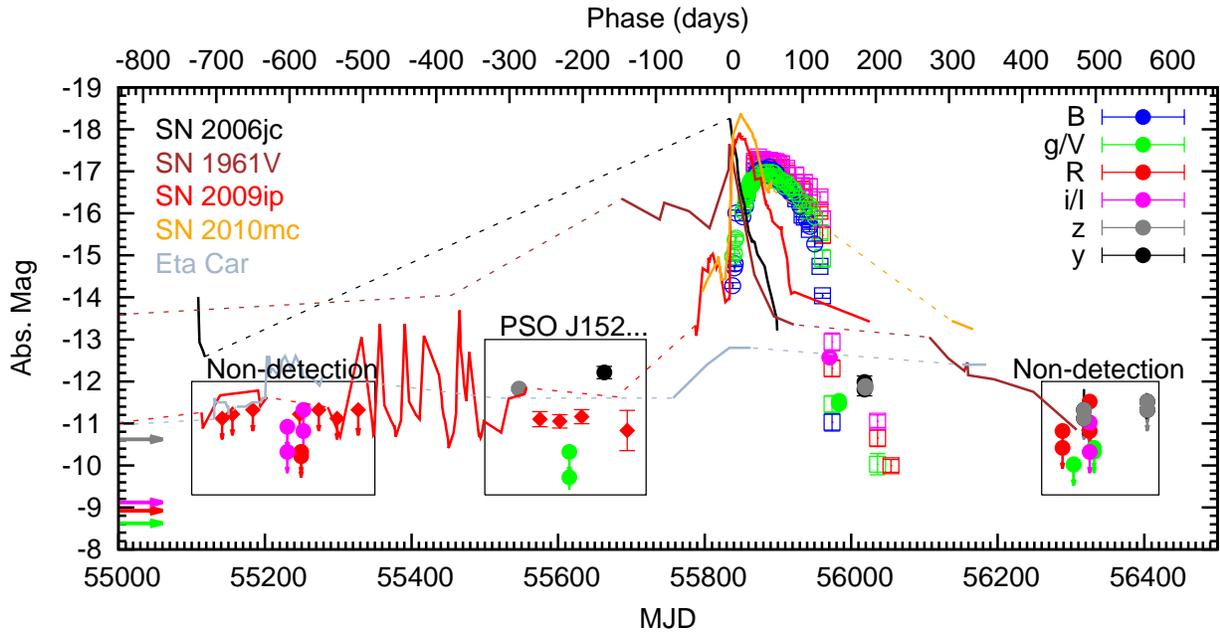}
\caption{
Observed lightcurve for SN 2011ht / PSO J152.0441+51.8492. PS1 photometry is marked with solid circles, while the values from \cite{Rom12} and \cite{Mau13b} are indicated with open circles and squares respectively. PS1 limits from non-detections are indicated with arrows with circles, while CRTS limits are shown with diamonds. The SDSS limits on the progenitor presented by \cite{Rom12} are marked with horizontal arrows on the left. Phase is with respect to the discovery epoch: MJD 55833.18. Also shown are the {\it R}-band lightcurves of five comparison objects which showed pre-explosion variability; the debated SN / SN impostor SN 1961V \citep{Zwi64}; SN 2009ip \citep{Fra13,Pas13}; SN 2010mc \citep{Ofe13}; the Ibn SN 2006jc \citep{Pas07} and the historical light curve of Eta Carinae around the Great Erruption of 1845 \citep{Smi11b}. The light curves of Eta Carinae, SNe 1961V and 2006jc are with respect to maximum brightness, SNe 2009ip and 2010mc are with respect to the estimated explosion epoch. Dashed lines connect poorly sampled sections of the lightcurves.
\label{fig:1}
}
\end{figure*}

The Catalina Real Time Transient Survey \citep[CRTS;][]{Dra09} covered the site of SN 2011ht with the 0.7 m Catalina Schmidt telescope (CSS) approximately once per month, for $\sim$4-6 months of each year from 2004 onwards. While the CRTS images are unfiltered, we have calibrated them to {\it r}$_\mathrm{SDSS}$, as this was the filter which showed the smallest scatter in the derived zeropoint. The explosion of SN 2011ht is clearly seen in the CRTS images taken on 2011 December 29. To search for a source in the preceding images,  we took the CRTS image from December 2005 as a template, stacked the four exposures from each night of CRTS observations between 2004 and the explosion of SN 2011ht, and performed difference imaging using the {\sc isis} package \citep{Ala00}. An example of the difference images obtained is shown in Fig.~\ref{fig:crts}. PSO~J152.0441+51.8492 was visible in the difference images taken between January and March 2011, with a subsequent tenuous detection in May 2011. PSF fitting photometry was performed on the difference images; the zeropoint was determined from aperture photometry of several sources in the field with SDSS {\it r} photometry.  The magnitudes of PSO~J152.0441+51.8492 are reported in Table \ref{table_phot}, and shown in Fig.~\ref{fig:1}. In other epochs where PSO J152.0441+51.8492 was not detected, a limiting magnitude at each was determined by adding and recovering artificial sources at the position of SN 2011ht, as shown in Fig.~\ref{fig:1}.

We also examined the series of images of UGC 5460 taken by one of us (T. B.) as part of a supernova search programme carried out over the period between April 1999 to September 2011. These images were taken every one to two months during the observing season for UGC 5460 (September - April) using three 0.36 m telescopes equipped with unfilltered, thinned, back illuminated AP7 CCD cameras. Aside from the discovery image of SN 2011ht \citep{Bol11}, no source was visible at the SN location. The limiting magnitude of these images is approximately {\it r}$\sim$18.5-19, and as these are less restrictive than the CRTS data, we do not consider them any further.

\section{Discussion}

A source, PSO J152.0441+51.8492, is clearly visible at the site of SN 2011ht between 287 d and 139 d prior to discovery. PSO J152.0441+51.8492 was not visible after SN 2011ht has faded at +484 days, nor when the field was observed again at +571 d, lending credence to the hypothesis that these sources are physically linked, and a variable unrelated background source is not the source of our findings. 

Unfortunately, we do not have images of PSO J152.0441+51.8492 taken on the same night with different filters. However, the non-detection of PSO J152.0441+51.8492 in \gps\ on Feb 23, coupled with the CRTS detections at a constant {\it r} magnitude both 13 days prior and 17 days later, points towards a red transient. Whether the red colour would be intrinsic or caused by dust is unclear, although both are plausible. For a $\sim$7000 K eruption, the {\it r-z} colour and non-detection in {\it g} are consistent with negligible extinction. We caution however, that SN 2009ip showed rapid variability (as can be seen in Fig.~\ref{fig:1} at $\sim$-450 days), and that we cannot hence exclude the possibility that  PSO J152.0441+51.8492 is varying by $\gtrsim$1 mag on a timescale of $\lesssim$2 weeks, and that this is the cause of the non-detection in \gps.

It is interesting to consider what the mechanism behind an eruption $\sim$1 year prior to a SN explosion could be. The Ne and O burning timescales for a massive star are on the order of months to a few years \citep{Woo02}, and \cite{Smi13b} has proposed that Ne flashes in 8-10 \msun\ progenitors could be a viable candidate mechanism. In this scenario, a temperature inversion develops in the semi-degenerate core of a star, due to the neutrino losses which are highest in the very centre. This temperature inversion can lead to off-centre Ne ignition \citep{Nom88}. As the core is partially supported by electron degeneracy, the Ne burning front which propagates inwards is unstable, and can give rise to explosive burning and possibly an ejection of part of the stellar envelope. However, in the models of \citeauthor{Nom88} this phenomenon occurs in only a small mass range, corresponding to a helium core of between 2.8-3.0 \msun. Furthermore, and as noted by \cite{Ume12}, this behaviour has not been probed with more modern codes, and hence it is uncertain if Ne flashes do in fact drive significant eruptions.

There are probably too many Type IIn SNe to all come from massive LBV-like progenitors \citep{Smi11c}. Indeed, given the heterogeneity of the class, it would seem natural to expect a similarly diverse range of progenitors. SN 2011ht has some similarity to SNe 1994W and 2009kn  \citep{Sol98,Kan12,Mau13b,Hum12,Rom12,Des09}, most notably an initial $\sim$120 d period of approximately constant luminosity, followed by a sudden drop of several magnitudes onto a tail phase, which led \cite{Mau13b} to propose these as a subclass of ``Type IIn-P'' SNe. Based on this, the faint tail, \citeauthor{Mau13b} suggested that these objects come from either low mass (8-10 \msun) progenitors with a weak explosion and low ejected $^{56}$Ni mass, or higher mass progenitors with fallback. The ``plateau'' in these objects is, however, perhaps distinct from the plateau in normal Type IIP SNe, as it is unclear whether they are dominated by the recombination of a H envelope \citep{Chu04}. PSO J152.0441+51.8492 provides the first evidence that, if they are indeed genuine core-collapse SNe, then at least one of this class of objects had an outburst $\lesssim$1 year prior to explosion.

Integrating under the light curve of PSO J152.0441+51.8492 gives a total energy of $\sim 10^{46}$ erg. This is admittedly only a crude estimate given the limited wavelength coverage of our data {\it grizy}, and that we implicitly assume that the colour of PSO J152.0441+51.8492 remained constant over the duration of the outburst. The true value is likely to be considerably higher. Nonetheless, for a characteristic CSM velocity of 600-700 \kms\ \citep{Rom12}, this corresponds to the kinetic energy of only 2$\times 10^{-3}$ \msun\ of ejecta. It is unclear whether such a small mass could provide a sufficiently dense CSM for strong interaction to occur one year later. This mass is also considerably lower than the $\sim$0.5 \msun\  of material ejected 1.5 yr prior to explosion which \cite{Chu04} proposed could explain the lightcurve of SN 1994W.  However, we again stress that the energy budget and variability history of PSO J152.0441+51.8492 are not well constrained, and hence the estimated mass could well be significantly higher. 

The faint, and perhaps red, nature of PSO J152.0441+51.8492 provides a tantalising hint that such outbursts may well be associated with other Type IIn SNe. PSO J152.0441+51.8492 was much fainter than both the first observed outburst of SN 2010mc, and the ``2012a'' eruption of SN 2009ip, but similar in magnitude to the earlier 2009-2011 eruptions of the latter. If such events are common, then with the ever-increasing cadence and sensitivity of all-sky surveys, many more of them will be discovered in future.

\acknowledgments

The Pan-STARRS1 Surveys (PS1) have been made possible through contributions of the Institute for Astronomy, the University of Hawaii, the Pan-STARRS Project Office, the Max-Planck Society and its participating institutes, the Max Planck Institute for Astronomy, Heidelberg and the Max Planck Institute for Extraterrestrial Physics, Garching, The Johns Hopkins University, Durham University, the University of Edinburgh, Queen's University Belfast, the Harvard-Smithsonian Center for Astrophysics, the Las Cumbres Observatory Global Telescope Network Incorporated, the National Central University of Taiwan, the Space Telescope Science Institute, the National Aeronautics and Space Administration under Grant No. NNX08AR22G issued through the Planetary Science Division of the NASA Science Mission Directorate, the National Science Foundation under Grant No. AST-1238877, and the University of Maryland. 

The CSS survey is funded by the National Aeronautics and Space Administration under Grant No. NNG05GF22G issued through the Science Mission Directorate Near-Earth Objects Observations Program.  The CRTS survey is supported by the U.S.~National Science Foundation under grants AST-0909182. This research has made use of the NASA/ IPAC Infrared Science Archive, which is operated by the Jet Propulsion Laboratory, California Institute of Technology, under contract with the National Aeronautics and Space Administration.

RK acknowledges funding from STFC. The research leading to these results has received funding from the European Research Council under the European Union's Seventh Framework Programme (FP7/2007-2013)/ERC Grant agreement n$^{\rm o}$ [291222]  (PI : S. J. Smartt). 

We thank Anders Jerkstrand for useful discussion on the late stages of stellar evolution, and the anonymous referee for a careful reading of the paper.

\clearpage

\end{document}